\documentclass[12pt]{article}

\usepackage{amssymb}
\usepackage{amsmath}
\usepackage{amsfonts}
\usepackage{color}
\usepackage{epsfig}
\usepackage{rotating}

\def\E {{\rm E}}

\newenvironment{proof}%
    {\par \noindent {\bf Proof}}%
    {\par \indent}

\newtheorem{theorem}{\bf Theorem}

\newtheorem{lemma}{\bf Lemma}

\newtheorem{remark}{\bf Remark}

\begin{document}

\title{{Nonparametric estimation of the entropy using a ranked set sample}}
\author{{{ Morteza Amini}$^{\dag}$}\footnote{Corresponding Author \newline {\it E-mail addresses:} morteza.amini@ut.ac.ir (Morteza Amini), mahdizadeh.m@live.com (Mahdi Mahdizadeh)}\, and {{ Mahdi Mahdizadeh}$^{\ddag}$}\\
{{$^\dag$ \small Department of Statistics, School of Mathematics, Statistics }}\\
{{\small and computer Science, College of Science, University of Tehran, }}\\
{{\small P.O. Box 14155-6455,  Tehran, Iran}}\\
{{$^\dag$ \small School of  Biological Sciences , Institute for Research in }}\\
{{\small  Fundamental Sciences (IPM), P. O.  Box 19395-5746, Tehran, Iran}}\\
{{$^\ddag$ \small Department of Statistics, Hakim Sabzevari University, P.O. Box 397, Sabzevar, Iran}}}
\maketitle
\begin{abstract} This paper is concerned with nonparametric estimation of the entropy in ranked set sampling.
Theoretical properties of the proposed estimator are studied. The
proposed estimator is compared with the rival estimator in simple
random sampling. The applications of the proposed estimator to the
mutual information estimation as well as estimation of the
Kullback-Leibler divergence are provided. Several Mont\'{e}-Carlo
simulation studies are conducted to examine the performance of the
estimator. The results are applied to the longleaf pine (Pinus
palustris) trees and the body fat percentage data sets to illustrate
applicability of theoretical results.
\end{abstract}


\section{Introduction}

In many fields of sciences, including environmental, biological and
ecological studies, measurement of the variable of interest is
usually time-consuming and/or costly. Under simple random sampling
(SRS), this leads to samples of small sizes with which statistical
inference is less reliable. Thus, if the situation permits, a more
efficient sampling technique may be desirable. A proper choice in
such situations is ranked set sampling (RSS) which was originally
introduced by McIntyre (1952) in the context of an agricultural
experiment. The application of RSS in many other areas of science is
studied by many researchers, including Kvam (2003), who applied RSS
for environmental monitoring, Halls and Dell (1966), who mentioned
the application of RSS in forestry  and Chen et al. (2005) who
applied RSS in medicine.

RSS is highly applicable when a few units in a set are easily ranked
without full measurement. One of the judgment ranking methods is to
use an auxiliary variable for ranking. The auxiliary variable, is
often highly correlated with the variable of interest and its
measurement is usually inexpensive and fast. For example, in a
clinical trial, the measurement of the body fat percentage variable
is time consuming and expensive. However, the measurement of the
body circumference (waist), which is highly correlated with the body
fat percentage, is inexpensive and fast. This auxiliary variable may
be used to extract a ranked set sample of the values of the variable
of interest. The judgment ranking is based on any means not
involving formal measurement, e.g. visual ranking or concomitant
variable.

Many authors have introduced extensions of RSS to construct improved estimators of different population attributes.
Ozturk (2011) proposed some variations on RSS in which rankers are permitted to declare ties.
 Multistage ranked set sampling (MRSS), proposed by Al-Saleh and Al-Omari (2002), is another
 example of such a design. The sampling methodology of MRSS is described in the following algorithm.
 Balanced RSS is simply the case $r=1$ of the following algorithm.
\begin{enumerate}
\item Randomly identify $k^{r+1}$, ($r\geq 1$) units from the target population and allocate them
randomly in $k^{r-1}$ sets, each of size $k^2$.

\item For each set in step 1, apply judgement ordering, by any cheap method not involving formal measurement, on the elements of the $i$th $(i=1,\ldots,k)$ sample and identify the $i$th (judgement) smallest unit, to get a (judgement) ranked set of size $k$. This step gives $k^{r-1}$ (judgement) ranked sets, each of size $k$.

\item Repeat step 2, $r-1$ times and apply each ranking stage on the ranked sets of its previous stage. A ranked set of size $k$ is acquired in the ($r-1$)th stage.

\item Actually measure the $k$ identified units in step 3.

\item Repeat steps 1-4, $m\geq 1$ times, to obtain a sample of size $n=mk$.
\end{enumerate}

We denote the sample collected using MRSS by
$\{X_{[i]j}:i=1\ldots,k;j=1,\ldots,m\}$, where $X_{[i]j}$ is the
$i$th judgement order statistic from the $j$th cycle. The special
case $r=2$ of MRSS (Al-Saleh and Al-Kadiri, 2000) is known as double
ranked set sampling (DRSS).

The ranked set sample contains not only measurements on the
quantified units but also additional information on their ranks.
Since RSS provides more structured samples than SRS, improved
inference may result from the use of RSS design.

Investigation of RSS in nonparametric settings has attracted attention of many researchers.
Dell and Clutter (1972) formally showed that the sample mean using RSS is an unbiased estimator of the population
mean regardless of ranking errors and it has a smaller variance than the sample mean using
SRS when the number of measured units are the same. Stokes (1980)
considered estimation of variance and showed that improved estimates of variance can be
produced in RSS. Stokes and Sager (1988) characterized a ranked set sample as a sample from a
conditional distribution, conditioning on a multinomial random vector, and applied RSS
to the estimation of the cumulative distribution function.
Chen (1999) studied the kernel method of density estimation in RSS.
Deshpande et al. (2006) have obtained nonparametric confidence intervals for
quantiles of the population based on a ranked set sample.

Since the concept of differential entropy was introduced in Shannon
(1968)'s paper, it has been of great interest because of wide
applications specially in signal analysis and analysis of neural
data and also because of its interesting theoretical properties. The
concept of mutual information, defined based on Shannon entropy, is
also well-known specially as a well-defined dependence measure (see
Cover and Thomas, 2012 for more details). The mutual information has
several advantages compared to the classical dependence measures,
such as Pearson, Spearman and Kendall's correlation. The classical
dependence criteria usually measure the correlation between two
variables. The Pearson correlation criterion only measures the
linear dependence and the Spearman and Kendall's correlation
criteria use only the ranks of the observations. The mutual
information enables us to measure whether linear or nonlinear
dependency between two or more variables.

Joe (1989) studied the large sample properties of the kernel based
estimator of the entropy based on SRS. He also obtained the optimal
bandwidth and a proper kernel function. In this paper, we are
specially interested in estimating the entropy based on RSS and MRSS
by adapting the work of Joe (1989). The proposed estimator is
applied to estimate the mutual information of variables as well as
the Kullback-Leibler divergence measure. We compare the proposed
kernel based estimator of the entropy based on RSS and MRSS samples
with that of Joe (1989) by approximating relative efficiency of the
proposed estimator with respect to its competitor in SRS. We observe
that the RSS and MRSS methods, perfectly improve the kernel based
estimation of the entropy.

The outline of this paper is as follows. Section 2 introduces a
nonparametric estimator of the entropy in RSS. The approximated bias
and mean square error (MSE) of the proposed estimator are derived.
The choice of kernel and the optimal bandwidth is discussed and
estimation of the MSE of the entropy estimator and approximating the
efficiency of the estimator relative to the corresponding estimator
in SRS are studied. The applications of the proposed estimator for
estimation of the mutual information and the Kullback-Leibler
divergence criteria is discussed in Section 3. A simulation study
and two real data examples are presented in Section 4 to examine
performance of the proposed estimators and to illustrate
applicability of theoretical results.
\section{The proposed estimator}

Let {$f_X$} be a $p$-variate {($p\geq 1$)} probability density function {(pdf)} with respect to Lebesgue measure $\mu$ with distribution function { $F_X$}.  We consider estimation of
$$H(X)=-\int_S f_X \log f_X\; {\rm d} \mu,$$
using a ranked set sample, where $S$ is a bounded set, such that
$f_X$ is bounded below on it by a positive constant and $\int_S f_X
\log f_X\; {\rm d} \mu\approx \int_{{\cal R}^p} f_X \log f_X\; {\rm
d} \mu$.

We assume that $f_X$ has continuous first and second derivatives,
which are dominated by integrable functions.

For each $i$, $X_{[i]1},\ldots,X_{[i]m}$ are independent and
identically distributed to $X_{[i]}$. For $p=1$ and the ordinary RSS
$(r=1)$, when the ranking is perfect, $X_{[i]}$ is identically
distributed to the $i$th order statistic in a sample of size $k$,
from the parent distribution $F$, while $X_{[1]},\ldots,X_{[k]}$ are
independent. For $p=2$, if the values of $X=(X^{(1)},X^{(2)})$ are
ranked by $X^{(1)}$, then the distribution of $X_{[i]}$ is identical
with the joint distribution of the $i$th order statistic and its
concomitant in an iid sample of size $k$ from $f_X$. For the perfect
ranking and DRSS case ($r=2$), the distribution of $X_{[i]}$ is
identical with that of the $i$th order statistic and its
concomitants of the independent but not identically distributed
sample $X_{[1]},\ldots,X_{[k]}$.

We assume throughout that the ranked set sample $X_{[i]j},\; i=1,\ldots,k,\;j=1,\ldots,m$ is consistent, that is
\begin{description}
\item (C1) For each $x$, \(\frac{1}{k}\sum_{i=1}^{k}F_{X_{[i]}}(x)=F_X(x).\)
\end{description}
It is well known that if the ranking of $X$ is performed using one
of its components, then both RSS and MRSS samples are consistent
(see e.g., Al-Saleh and Al-Omari (2002) for a proof of the
consistency for the DRSS case).

For any $t$ in support of $X$, the kernel density estimate of $f_X(t)$ based on the ranked set sample $X_{[i]j},\; i=1,\ldots,k,\; j=1,\ldots,m$, is (Chen, 1999)
\[f_{n}^{\rm RSS}(t)=\frac{1}{n\gamma^p}\sum_{i=1}^{k}\sum_{j=1}^{m}{K_p}\left(\frac{t-X_{[i]j}}{{\gamma}}\right),\]
where  $K_p$ is a kernel function and {$\gamma$} is a positive bandwidth. We assume throughout that {$K_p(u_1,\ldots,u_p)=\prod_{j=1}^p k_0(u_j)$ and $k_0$ is a univariate kernel function satisfying the following conditions:
\begin{description}
\item (C2) $k_0(u)=k_0(-u)$;\hspace{0.5cm}(C3) $\int k_0(u) \; {\rm d} u =1$ \hspace{0.25cm}and \hspace{0.25cm} (C4) $\int u^2 k_0(u) \; {\rm d}u=1$.
\end{description}}
Let
\[F_{n}^{\rm RSS}(t)=\frac{1}{n}\sum_{i=1}^{k}\sum_{j=1}^{m}I(X_{[i]j}\leq t)\]
be the estimator of $F$ in RSS, then
$$f_{n}^{\rm RSS}(t)=\int \frac{1}{\gamma^p}K_p\left(\frac{t-u}{\gamma}\right)\; {\rm d}F_{n}^{\rm RSS}(u).$$
We propose the estimator of the entropy of $X$, $H(X)$, in RSS based on $X_{[i]j},\; i=1,\ldots,k,\; j=1,\ldots,m$, as
\[H_{n}^{\rm RSS}(X)=-\frac{1}{n}\sum_{i=1}^{k}\sum_{j=1}^{m}\log f_{n}^{\rm RSS}(X_{[i]j})I_{S}(X_{[i]j})=-\int_{S} \log f_{n}^{\rm RSS}(x)\; {\rm d} F_{n}^{\rm RSS}(x).\]

The asymptotic properties of $H_{n}^{{\rm SRS}}(X)$ are studied by
Joe (1989). In the next section, we study the asymptotic properties
of $H_{n}^{{\rm RSS}}(X)$.

\subsection{Characteristics of the estimator}
{In this section, we {approximate} MSE and bias of $H_n^{\rm RSS}(X)$. With a suitable choice of bandwidth $\gamma$ (see Section 4), we deduce that MSE($H_n^{\rm RSS}(X)$) is $O(n^{-1})$, that is $H_n^{\rm RSS}(X)$ is a consistent estimator of $H(X)$. To prove the results, we need the following lemma, {which is an analog of Lemma 2.1 of Joe (1989).}
\begin{lemma}\label{l1}
Let $U_{n}(x)=\sqrt{n}(F_{n}^{\rm RSS}(t)-F(t))$, then, {under the condition {C1},}
\begin{description}
\item (i) for any integrable function $a(x)$ with $\E(a(X))<\infty$, we have $\E\int a(x)\; {\rm d} U_{n}(x)=0;$
\item (ii) for any integrable function $a(x,y)$, we have
\begin{align*}
\E\int\int a(x,y)\; {\rm d} U_{n}(x)\; {\rm d} U_{n}(y)&=\int a(x,x)\; {\rm d} F_X(x)\\
&-\frac{1}{k}\sum_{i=1}^{k}\int\int a(x,y)f_{X_{[i]}}(x)f_{X_{[i]}}(y)\; {\rm d} x\; {\rm d} y.
\end{align*}
\end{description}
\end{lemma}
\begin{proof}
See the Appendix.
\end{proof}
By using Lemma \ref{l1}, we get the following result.
\begin{theorem} \label{t1}
\begin{description}
\item (i) The bias of $H_{n}^{\rm RSS}(X)$ is $$\E(H_{n}^{\rm RSS}(X))-H(X)=(H_{\gamma}-H(X))+n^{-1}\alpha_1+O(n^{-3/2}),$$
where $H_{\gamma}=-\E(I_{S}(X)\log(\Delta_{\gamma}(X)))$, $\Delta_{\gamma}(x)=\E\left(\frac{1}{\gamma^p}K\left(\frac{x-X}{\gamma}\right)I_{S}(X)\right)$,
$${\alpha_1=\int_{S} {\cal B}(x,x)\; {\rm d} F_X(x)-\frac{1}{k}\sum_{i=1}^{k}\int_{S}\int_{S} {\cal B}(x,y)f_{X_{[i]}}(x)f_{X_{[i]}}(y)\; {\rm d} x\; {\rm d} y,}$$
and
\begin{equation}\label{Bn}
{\cal B}(x,y) =\frac{-1}{2}\int_{S}\frac{\frac{1}{\gamma^{2p}}K_p\left(\frac{z-x}{\gamma}\right)K_p\left(\frac{z-y}{\gamma}\right)}{(\Delta_{\gamma}(z))^2}\;{\rm d} F_X(z)+\frac{\frac{1}{\gamma^p}K_p\left(\frac{y-x}{\gamma}\right)}{\Delta_{\gamma}(x)}I_{S}(x).
\end{equation}
\item (ii) The MSE of $H_{n}^{\rm RSS}(X)$ is
\begin{equation}\label{mse}
\E(H_{n}^{\rm RSS}(X)-H(X))^2=(H_{\gamma}-H(X))^2+n^{-1}(\alpha_2-2\alpha_1(H_{\gamma}-H(X)))+O(n^{-3/2}),
\end{equation}
where
$${\alpha_2=\int_{S} {\cal A}^2(x)\; {\rm d} F_X(x)-\frac{1}{k}\sum_{i=1}^{k}\left(\int_{S}{\cal A}(x)\; {\rm d} F_{X_{[i]}}(x)\right)^2,}$$
and
\begin{equation}\label{An}
{\cal A}(x)=\int_{S}\frac{\frac{1}{\gamma^p}K_p\left(\frac{y-x}{\gamma}\right)}{\Delta_{\gamma}(y)}\;{\rm d} F_X(y)+\log(\Delta_{\gamma}(x))I_{S}(x)
\end{equation}

\end{description}
\end{theorem}
\begin{proof}
See the Appendix.
\end{proof}

\subsection{Choosing the optimal bandwidth and kernel}

To choose an optimal value for the bandwidth and a suitable kernel function, note that
\begin{align*}
\alpha_1&=\theta+\gamma^{-p}K(0)\int_{S}\Delta_{\gamma}(x)^{-1}\;{\rm d} F_X(x)-\gamma^{-p}\frac{\kappa_2}{2}\int_{S}\Delta^*_{\gamma}(z)\Delta_{\gamma}(z)^{-2}\;{\rm d} F_X(z)\\
&=\theta-\gamma^{-p}\left(\frac{\kappa_2}{2}-K(0)\right)\int_{S}\Delta_{\gamma}(x)^{-1}\;{\rm d} F_X(x)\\
&-\gamma^{-p}\frac{\kappa_2}{2}\int_{S}(\Delta^*_{\gamma}(z)-\Delta_{\gamma}(z))\Delta_{\gamma}(z)^{-2}\;{\rm d} F_X(z),
\end{align*}
where $$\theta=-\frac{1}{k}\sum_{i=1}^{k}\int_{S}\int_{S} {\cal B}(x,y)f_{X_{[i]}}(x)f_{X_{[i]}}(y)\; {\rm d} x\; {\rm d} y,$$
$\Delta^*_{\gamma}(z)=\E\left(I_{S}(X)\gamma^{-p}K^2((z-X)/\gamma)\right)/\kappa_2$ and $\kappa_2=\int K^2(u)\;{\rm d} u$.

The term $\theta$ is $O(1)$. In addition,
\begin{align*}
\Delta^*_{\gamma}(z)-\Delta_{\gamma}(z)=(2\gamma^2)^{-1}{\rm tr }f''_X(z)\left[\int \frac{v^2k_0^2(v)}{\kappa_{02}}\;{\rm d}v-1\right]+O(\gamma^2),
\end{align*}
where $\kappa_{02}=\int k_0^2(u)\;{\rm d} u$.

Thus, if
\begin{equation}\label{k01}
k_0(0)=\frac{\kappa_{02}}{2^{1/p}}
\end{equation}
and
\begin{equation}\label{k02}
\int \frac{v^2k_0^2(v)}{\kappa_{02}}\;{\rm d}v=1,
\end{equation}
then, we have
$$\alpha_1=O(1)+O(\gamma^{2-p}).$$
Joe (1989) showed that the kernel function of the form
\begin{equation}\label{kjoe}
k_0(u)=\left\{\begin{array}{l l}\eta_1+\eta_1|u|,& |u|<\xi_1\\
\eta_3-\eta_4|u|,&\xi_1<|u|<\xi_2.\end{array}\right.
\end{equation}
satisfies conditions \eqref{k01} and \eqref{k02} and computed $\eta_i,\xi_j,\;i=1,\ldots,4,\; j=1,2$ for $p=1,\ldots,4$.

For $p\leq 2$, a competitor of the kernel function in \eqref{kjoe} would be
\begin{equation}\label{knorm}
k_0(u)=(4\pi)^{-0.5}\exp\{-u^2/4\},
\end{equation}
which satisfies condition \eqref{k02}.

Using \eqref{knorm}, we have, for $p\leq 2$
$$\alpha_1=O(1)+O(\gamma^{-p}).$$

On the other hand, since $H_{\gamma}-H(X)=O(\gamma^2)$, for the kernel function in \eqref{kjoe}, we get
\[\E(H_{n}^{\rm RSS}(X)-H(X))^2=O(n^{-1})+O(n^{-1}\gamma^2)+O(n^{-1}\gamma^{4-p})+O(\gamma^4)\]
and
\[\E(H_{n}^{\rm RSS}(X)-H(X))=O(n^{-1})+O(n^{-1}\gamma^{2-p})+O(\gamma^2),\]
and for the kernel function in \eqref{knorm}, we get
\[\E(H_{n}^{\rm RSS}(X)-H(X))^2=O(n^{-1})+O(n^{-1}\gamma^2)+O(n^{-1}\gamma^{2-p})+O(\gamma^4)\]
and
\[\E(H_{n}^{\rm RSS}(X)-H(X))=O(n^{-1})+O(n^{-1}\gamma^{-p})+O(\gamma^2).\]

Using the kernel function in \eqref{kjoe}, for $p\leq 2$, by setting
$\gamma=O(n^{-1/(2+0.5p)})$, we deduce that
\[\E(H_{n}^{\rm RSS}(X)-H(X))^2=O(n^{-1})+O(n^{-4/(2+0.5p)})\]
and
\[\E(H_{n}^{\rm RSS}(X)-H(X))=O(n^{-2/(2+0.5p)}).\]

Using the kernel function \eqref{knorm}, for $p\leq 2$, by setting $\gamma=O(n^{-1/(2+0.5p)})$, we have
\[\E(H_{n}^{\rm RSS}(X)-H(X))^2=O(n^{-1})+O(n^{-(4-0.5p)/(2+0.5p)})\]
and
\[\E(H_{n}^{\rm RSS}(X)-H(X))=O(n^{-(2-0.5p)/(2+0.5p)}).\]

Thus, using both kernels in \eqref{kjoe} and \eqref{knorm}, for $p\leq 2$, $\E(H_{n}^{\rm RSS}(X)-H(X))^2=O(n^{-1})$. Although the rate of convergence of the bias is somehow higher using the kernel in \eqref{kjoe}, simulations show that, for $p\leq 2$, using the kernel function in \eqref{knorm}, results in better performance of the entropy estimator than using the kernel function in \eqref{kjoe}, specially for smaller sample sizes.

From the above discussion, for both kernel functions in \eqref{knorm} and \eqref{kjoe} and for $p\leq 2$, the optimal bandwidth is of the form
$$\gamma=c\; n^{\frac{-1}{2+0.5p}},$$
where $c$ depends on the parameters of $f_X$.

To discuss about the suitable value of $c$, first, we study the behavior of the optimal value of $\gamma$ through simulation, obtained by minimizing the cross-validation criterion
\begin{equation}\label{Cg}
CV_{\gamma}=\frac{1}{m}\sum_{j=1}^m\left(H_n^{\rm RSS}(X)-H_n^{\rm RSS}(X^{(-j)})\right)^2,
\end{equation}
where $X^{(-j)}=\{X_{[i]l},\; i=1,\ldots,k,\; l(\ne j)=1,\ldots,m\}$. Simulations show that the bandwidth should decrease with more dependence for a bivariate density. Therefore, if $\Sigma$ is the covariance matrix of the population, then $c$ increases by increasing $\theta=\frac{|\Sigma|^{\frac{1}{2}}}{({\rm tr}(\Sigma^{-1}))^2}$. For the multivariate normal distribution with covariance matrix $\Sigma$ (with $|\Sigma|$ not too close to zero), a rough approximation of $\theta^{\frac{1}{2+0.5p}}$ is $\frac{0.5-\int_R \phi(x,\Sigma){\rm d}x}{p^{\frac{2}{2+0.5p}}(0.5-0.5^p)}$, where $R=[-0.674,0.674]^p$ (see Jones, 1989). Thus, we take the the optimal bandwidth to be
$$\gamma=d_1\; n^{\frac{-1}{2+0.5p}}\overline{\rm IQR}\frac{0.5-\hat{\alpha}}{0.5-0.5^p},$$
where $d_1$ is a suitable normalizing constant, which is decreasing in $p$, $\hat{\alpha}$ is the proportion of data in the rectangle $\times_{j=1}^p[q_{j1},q_{j3}]$, $q_{j1}$ and $q_{j3}$ are the lower and upper quartiles and $\overline{\rm IQR}$ is the average of interquartile ranges.

Simulations suggest that for $p \leq 2$,
the rule
\begin{equation}\label{rule}
\gamma=d_1\; n^{\frac{-1}{2+0.5p}}\overline{\rm IQR}\frac{0.5-\hat{\alpha}}{0.5-0.5^p},
\end{equation}
works quite well, with the kernel in \eqref{knorm}, for
$\rho=0.5(0.1)0.9$, for the values of $d_1$, given in Table
\ref{tr}, for different values of $k=3,5$, $p=1,2$ and $r=1,2$. A
value of $d_1$ which was corresponding to the minimum value of the
simulated MSE of the estimator $H_n^{\rm RSS}(X)$ on a grid of
values of $d_1=0.5(0.05)2.00$ is chosen as the optimal value of
$d_1$.

\begin{remark}
Simulations suggest that for $p=3,4$, the bandwidth in \eqref{rule}
works quite well with the kernel in \eqref{kjoe} under the
multivariate normal distribution. We have not provided the suitable
values of $d_1$ for this case here for the sake of briefness.
However, an illustrative example is provided for the case $p=3$ in
Section 4.
\end{remark}

\begin{table}
\centering
\caption{The suitable choices of $d_1$ in \eqref{rule} for the estimator of $H_n(X)$ under the bivariate normal distribution, for $n=30$, $k=3,5$, $p=1,2$, $r=1,2$ and $\rho=0.5(0.1)0.9$.}\label{tr}
\begin{tabular}{c c c | c c c c c}
\hline\hline
       &     &    &       &       &  $\rho$ &   &   \\
$r$ & $k$ & $p$ & 0.5 & 0.6 & 0.7 & 0.8 & 0.9 \\
\hline
  1   &   3   & 1 & 1.45 & 1.40 & 1.30 & 1.25 & 1.20 \\
       &       & 2 & 1.45 & 1.45 & 1.30 & 1.30 & 1.05\\
        &   5  &1  & 1.45 &  1.40 & 1.30 & 1.25 &1.20\\
        &      & 2 & 1.50 & 1.45 &1.45  &1.30  & 1.05\\
   2   &   3  &1  &1.50 & 1.45 & 1.45  &1.40  & 1.20\\
          &     &2  & 1.45 & 1.45 & 1.40  &1.30  & 1.05\\
          &   5  & 1 &  1.50 & 1.45 & 1.45 &1.40 &1.30 \\
          &      & 2 & 1.50 & 1.45 & 1.40 & 1.20 & 1.05\\
\hline\hline
\end{tabular}
\end{table}


\subsection{Estimation of the MSE of the estimator}

Computations show that $CV_{\gamma}$ in \eqref{Cg} has a large negative bias for estimating MSE of $H_{n}^{\rm RSS}(X)$. To propose a less biased estimator for MSE of $H_{n}^{\rm RSS}(X)$, we use the approximated MSE in Theorem \ref{t1}, where $CV_{\gamma}$ is employed as an estimator of the term
$({H}_{\gamma}-H(X))^2$. Since ${H}_{\gamma}-H(X)$ is negative for suitable values of $\gamma$, we use $-|D_{\gamma}|$ as an estimator of ${H}_{\gamma}-H(X)$, where
$$D_{\gamma}=\frac{1}{m}\sum_{j=1}^m\left(H_n^{\rm RSS}(X)-H_n^{\rm RSS}(X^{(-j)})\right).$$
Thus, our proposed estimator of the MSE of $H_{n}^{\rm RSS}(X)$ is
\begin{equation}\label{Mg}
\hat{M}_{\gamma}=CV_\gamma+n^{-1}(\hat{\alpha}_2+2\hat{\alpha}_1|D_{\gamma}|)\delta(CV_\gamma+n^{-1}(\hat{\alpha}_2+2\hat{\alpha}_1|D_{\gamma}|)),
\end{equation}
where
$$\delta(t)=\left\{\begin{array}{l r}1& t\geq 0\\ 0 & t<0\end{array}\right .,$$
$$\hat{\alpha}_1=\frac{1}{mk}\sum_{i=1}^{k}\sum_{j=1}^{m}\hat{\cal B}(X_{[i]j},X_{[i]j})I_{S}(X_{[i]j})-
\frac{1}{m(m-1)k}\sum_{i=1}^{k}\sum_{j\neq j'(=1)}^{m}\hat{\cal B}(X_{[i]j},X_{[i]j'})I_{S}(X_{[i]j})I_{S}(X_{[i]j'}),$$
$$\hat{\alpha}_2=\frac{1}{mk}\sum_{i=1}^{k}\sum_{j=1}^{m}\hat{\cal A}^2(X_{[i]j})I_{S}(X_{[i]j})-
\frac{1}{k}\sum_{i=1}^{k}\left[\sum_{j=1}^{m}\hat{\cal A}(X_{[i]j})I_{S}(X_{[i]j})\right]^2,$$
$$\hat{\cal B}(x,y)=\frac{-1}{2mk}\sum_{i=1}^{k}\sum_{j=1}^{m}\frac{\frac{1}{\gamma^{2p}}K_p\left(\frac{X_{[i]j}-x}{\gamma}\right)K_p\left(\frac{X_{[i]j}-y}{\gamma}\right)}{(f_n^{\rm RSS}(X_{[i]j}))^2}I_{S}(X_{[i]j})+
\frac{\frac{1}{\gamma^p}K_p\left(\frac{y-x}{\gamma}\right)}{f_n^{\rm RSS}(x)}I_{S}(x)$$
and
$$\hat{\cal A}(x)=\frac{1}{mk}\sum_{i=1}^{k}\sum_{j=1}^{m}\frac{\frac{1}{\gamma^p}K_p\left(\frac{X_{[i]j}-x}{\gamma}\right)}{f_n^{\rm RSS}(X_{[i]j})}I_{S}(X_{[i]j})+\log(f_n^{\rm RSS}(x))I_{S}(x).$$

The performance of the estimators in \eqref{Cg} and \eqref{Mg} are
examined through simulation studies in Section 4.

\subsection{The relative efficiency }

To approximate  RE of $H_{n}^{\rm RSS}(X)$ with respect to $H_{n}^{\rm SRS}(X)$, we use the {approximated} MSE of $H_{n}^{\rm SRS}(X)$ obtained by Joe (1989) as
\begin{equation}\label{joe}
\E(H_{n}^{\rm SRS}(X)-H(X))^2=(H_{\gamma}-H(X))^2+n^{-1}(\beta_2-2\beta_1(H_{\gamma}-H(X)))+O(n^{-3/2}),
\end{equation}
where
\begin{align*}
\beta_1&=\int {\cal B}(x,x)\; {\rm d} F_X(x)-\int\int {\cal B}(x,y)\; {\rm d} F(x)\; {\rm d} F_X(y)\\
&=\int {\cal B}(x,x)\; {\rm d} F_X(x)-\frac{1}{2},
\end{align*}
and
\begin{align*}
\beta_2&=\int {\cal A}^2(x)\; {\rm d} F_X(x)-\left(\int{\cal A}(x)\; {\rm d} F_X(x)\right)^2\\
&=\int {\cal A}^2(x)\; {\rm d} F_X(x)-(1-H_{\gamma})^2,
\end{align*}
in which ${\cal A}(x)$ and $ {\cal B}(x,y)$ are given in \eqref{An} and \eqref{Bn}, respectively. Joe (1989) also showed that
the bias of $H_{n}^{\rm SRS}(X)$ is
$$\E(H_{n}^{\rm SRS}(X))-H(X)=(H_{\gamma}-H(X))+n^{-1}\beta_1+O(n^{-3/2}).$$

As a corollary of Theorem \ref{t1} and using \eqref{joe}, the RE of $H_{n}^{\rm RSS}(X)$ relative to $H_{n}^{\rm SRS}(X)$ is approximated as
\begin{equation}\label{are}
{{\rm RE}(H_{n}^{\rm RSS}(X),H_{n}^{\rm SRS}(X))\approx}1+\frac{n^{-1}(\beta_2-\alpha_2-2(\beta_1-\alpha_1)(H_{\gamma}-H(X)))}{(H_{\gamma}-H(X))^2+n^{-1}(\alpha_2-2\alpha_1(H_{\gamma}-H(X)))}.
\end{equation}

Using Cauchy-Schwartz inequality for series and consistency of the RSS sample we have
\begin{align*}
\frac{1}{k}\sum_{i=1}^{k}\left(\int_{S}{\cal A}_{n}(x)\; {\rm
d} F_{X_{[i]}}(x)\right)^2&\geq
\left(\int_{A_{n}}{\cal A}_{n}(x)\; {\rm d} F_X(x)\right)^2
\end{align*}
and consequently $\beta_2\geq\alpha_2$.

Numerical computations under the bivariate normal distribution, for different values of $k$, $m$ and $\rho$, show that that for a suitable choice of $\gamma$, the approximated relative efficiency in \eqref{are} is greater than 1.
The computed values of the approximated relative efficiencies of $H(X^{(1)})$ are given in Table \ref{ret}, for different values of $\rho$, $n$ and $k$, for RSS and DRSS schemes, under bivariate normal distribution.

\begin{table}
\centering
\small\caption{Approximated relative efficiencies, under bivariate normal distribution, for different values of $\rho$, $n$ and $k$.\label{ret}}
\begin{tabular}{c | c c c c c c c c c c c c }\hline\hline
{$\rho$}&  0.9 &    &          &    &           &    &   0.8 &    &          &    &           &    \\
{$n$}   & 15    &    & 30     &    &  45     &    & 15    &    & 30     &    &  45     &    \\
{$k$}   & 3      & 5 &  3      & 5 &  3      & 5   & 3      & 5 &  3      & 5 &  3      & 5   \\
\hline
RSS   &   1.10       &  1.58   &     1.05  &  1.33   &    1.02     &   1.24      &   1.04         &  1.31    &    1.02    &   1.16 &     1.02    &  1.21  \\
DRSS &     1.68    &   2.04  &    1.29      &  1.44   &    1.23       &    1.28   &     1.32     &  1.50  &    1.25      &  1.32  &        1.18   & 1.25  \\
\hline
\end{tabular}
\end{table}

Since the values given in Table \ref{ret} are population parameters, different optimal values of $\gamma$ are needed for this computation. We used $\gamma=cn^{-1/(n+0.5p)}$, with suitable values of the normalizing constant $c$. It is found that, for the SRS scheme, suitable values of $c$ are $c=1.35$ for $\rho=0.9$ and $c=1.30$ for $\rho=0.8$. For RSS, we used $c=1.40$ for $k=3$ and $\rho=0.9$, $c=1.35$ for $k=3$ and $\rho=0.8$, $c=1.65$ for $k=5$ and $\rho=0.9$ and $c=1.60$ for $k=5$ and $\rho=0.8$. For DRSS, our suggested values are $c=1.45$ for $k=3$ and $\rho=0.9$, $c=1.40$ for $k=3$ and $\rho=0.8$, $c=1.70$ for $k=5$ and $\rho=0.9$ and $c=1.65$ for $k=5$ and $\rho=0.8$.

As one can see from Table \ref{ret}, all approximated values of ${\rm RE}(H_{n}^{\rm RSS}(X^{(1)}),H_n^{\rm SRS}(X^{(1)}))$ are greater than 1.00. Also, the relative efficiencies decrease as $\rho$ and/or $n$ increase and also as $k$ decreases. For DRSS, the relative efficiencies are greater than for RSS, as expected.

Finally, using the results of Al-Saleh and Al-Omari (2002) we have, for $i=1,\ldots,k$ and $x$  in the support of $X$,
\begin{equation}\label{Fiinf}
\lim_{r\to\infty}f_{X_{[i]}}(x)=\left\{\begin{array}{l l}
kf_X(x),& F_X^{-1}\left(\frac{i-1}{k}\right)\leq x<F_X^{-1}\left(\frac{i}{k}\right)\\
0, & \mbox{otherwise},
\end{array}\right.
\end{equation}
and consequently for each $x,y$  in the support of $X$,
\[\lim_{r\to\infty}\frac{1}{k}\sum_{i=1}^k f_{X_{[i]}}(x)f_{X_{[i]}}(y)=kf_X(x)f_X(y).\]
Thus
\[\lim_{r\to\infty}\alpha_1=\beta_1-\frac{k-1}{2}, \quad \mbox{and}\quad \lim_{r\to\infty}\alpha_2=\beta_2-(k-1)(1-H_{\gamma})^2,\]
and thereby
\begin{align*}
\lim_{r\to\infty}{\rm RE}(H_{n}^{\rm RSS}(X),H_{n}^{\rm SRS}(X))&\approx 1+n^{-1}(k-1)((1-H_{\gamma})^2-H_{\gamma}+H(X))\\
&\times \left[(H_{\gamma}-H(X))^2+n^{-1}(\beta_2-(k-1)(1-H_{\gamma})^2\right.\\
&\left.-2(\beta_1-\frac{k-1}{2})(H_{\gamma}-H(X)))\right]^{-1}.
\end{align*}

\section{Applications}
In addition to estimation of the entropy, the developed results
might be applied for estimation of the Kullback-Leibler distance
measure as well as the mutual information criterion. The estimator
of the Kullback-Leibler distance measure can be used to construct
homogeneity test statistics. The mutual information is well-known as
a measure of dependence. These applications are described in the
following sections.
\subsection{Application to estimation of the mutual information}

For $p\geq 2$ and $X=(X^{(1)}, X^{(2)})$, the mutual information of $X^{(1)}$ and $X^{(2)}$, defined as
\begin{align*}
I(X^{(1)},X^{(2)})&=H(X^{(1)})+H(X^{(2)})-H(X)\\
&=\int_S f_X \log \left(\frac{f_X}{f_{X^{(1)}}f_{X^{(2)}}}\right)\;
{\rm d} \mu,
\end{align*}
which is the Kullback-Leibler divergence between the joint
distribution of $X=(X^{(1)}, X^{(2)})$ and the product of its
marginal pdfs, is used as a popular and well-defined  nonlinear
correlation criteria.

If for example, $X$ is distributed as the bivariate normal distribution with correlation parameter $\rho$, the mutual information between $X^{(1)}$ and $X^{(2)}$ is
$$I(X^{(1)},X^{(2)})=-0.5\log(1-\rho^2).$$

Motivated by the above relationship, and using the fact that $I(X^{(1)},X^{(2)})\geq 0$, a standardized nonlinear correlation measure based on the mutual information might be defined as follows
\begin{equation}\label{StI}
{\cal I}(X^{(1)},X^{(2)})=1-e^{-2I(X^{(1)},X^{(2)})}\in(0,1).
\end{equation}

Using the proposed estimator of the entropy, the mutual information $I(X^{(1)},X^{(2)})$
is estimated by
\begin{align}
\hat{I}(X^{(1)},X^{(2)})&=H_n^{\rm RSS}(X^{(1)})+H_n^{\rm RSS}(X^{(2)})-H_n^{\rm RSS}(X)\nonumber\\
&=\int_S \log \left(\frac{f_n^{X;{\rm
RSS}}(x_1,x_2)}{f_n^{X^{(1)};{\rm RSS}}(x_1)f_n^{X^{(2)};{\rm
RSS}}(x_2)}\right)\; {\rm d} F_n^{X;{\rm RSS}}(x_1,x_2),\label{Ihat}
\end{align}
where $H_n^{{\rm RSS}}(X^{(\ell)})$ is the the estimator of $H(X^{(\ell)})$ in RSS based on $X^{(\ell)}_{[i]j}, \; i=1,\ldots,k,\; j=1,\ldots,m$, $\ell=1,2$ and $H_{n}^{{\rm RSS}}(X)$ is the estimator of $H(X)$ based on $(X^{(1)}_{(i)j},X^{(2)}_{[i]j}), \; i=1,\ldots,k,\; j=1,\ldots,m$, that is
$$H_{n}^{\rm RSS}(X)=-\frac{1}{mk}\sum_{i=1}^{k}\sum_{j=1}^{m}\log f_{n}^{\rm RSS}(X^{(1)}_{(i)j},X^{(2)}_{[i]j})I_{S}(X^{(1)}_{(i)j},X^{(2)}_{[i]j}).$$

Using the kernel function $K_p(u_1,\ldots,u_p)=\prod_{j=1}^p
k_0(u_j)$ and a similar bandwidth parameter for estimation of
$f_n^{X;{\rm RSS}}(x_1,x_2)$, $f_n^{X^{(1)};{\rm RSS}}(x_1)$ and
$f_n^{X^{(2)};{\rm RSS}}(x_2)$, it is straightforward that
$\hat{I}(X^{(1)},X^{(2)})\geq 0$. The corresponding estimator of
${\cal I}(X^{(1)},X^{(2)})$ is obtained by plugging in the estimator
of $\hat{I}(X^{(1)},X^{(2)})$ in \eqref{StI}.

Simulations show that the bandwidth parameter in \eqref{rule} works
well for the estimator in \eqref{Ihat}, with $p$ equal to dimension
of $X=(X^{(1)},X^{(2)})$, by choosing some suitable value of the
constant $d_1$. Table \ref{idt} presents the suitable choices of
$d_1$ in \eqref{rule}, for the estimator in \eqref{Ihat}, under the
bivariate normal distribution, for $n=30$, $p=2$ and different
values of $k$, $r$ and $\rho=0.5(0.1)0.9$. A value of $d_1$ which
was corresponding to the minimum value of the simulated MSE of the
estimator in \eqref{Ihat}, on a grid of values of
$d_1=0.5(0.05)2.00$, is chosen as the optimal value of $d_1$.

\begin{table}
\centering
\caption{The suitable choices of $d_1$ in \eqref{rule} for the estimator in \eqref{Ihat} under the bivariate normal distribution, for $n=30$, $p=2$ and different values of $k$, $r$ and $\rho$.}\label{idt}
\begin{tabular}{c c | c c c c c}
\hline\hline
       &        &       &       &  $\rho$ &   &   \\
$r$ & $k$ & 0.5 & 0.6 & 0.7 & 0.8 & 0.9 \\
\hline
  1   &   3   & 1.55 & 1.40 & 1.30 & 1.00 & 0.70\\
        &   5   & 1.50 & 1.30 & 1.30 & 1.00 & 0.70\\
   2   &   3   & 1.50 & 1.40 & 1.10 & 1.00 & 0.70\\
        &   5   &  1.50 & 1.40 & 1.30 & 1.10 & 1.00\\
\hline\hline
\end{tabular}
\end{table}

\subsection{Application to estimation of the Kullback-Leibler divergence}

Using the proposed estimator of the entropy, the Kullback-Leibler divergence between $f_{X^{(1)}}$ and $f_{X^{(2)}}$ defined as
\begin{equation}\label{KL}
KL(f_{X^{(1)}},f_{X^{(2)}})=\int_S f_{X^{(1)}} \log \left(\frac{f_{X^{(1)}}}{f_{X^{(2)}}}\right)\; {\rm d} \mu,
\end{equation}
can be estimated as
\begin{eqnarray}
\widehat{KL}(f_{X^{(1)}},f_{X^{(2)}})&=&\int_S \log
\left(\frac{f_n^{X^{(1)};{\rm RSS}}(x)}{f_n^{X^{(2)};{\rm
RSS}}(x)}\right)\; {\rm d} F_n^{X^{(1)};{\rm
RSS}}(x)\nonumber\\
&=&\frac{1}{mk}\sum_{i=1}^{k}\sum_{j=1}^{m}\log
\left(\frac{f_n^{X^{(1)};RSS}(X^{(1)}_{[i]j})}{f_n^{X^{(2)};RSS}(X^{(1)}_{[i]j})}\right).\label{klh}
\end{eqnarray}

Statistical properties of the estimator in \eqref{klh} as well as
its application for the homogeneity test remains as an open problem.

\section{Numerical studies}
In this section, several numerical studies are conducted to examine
the performance of the proposed estimators and to illustrate
theoretical results. A simulation study under the bivariate normal
distribution is conducted to examine the performance of the
estimators in \eqref{Cg} and \eqref{Mg}. A data analysis is
performed to examine the performance of the estimators for a real
data example. Finally, the  proposed theoretical results are applied
to the problem of variable selection based on the estimator of the
mutual information as a correlation measure.
\begin{sidewaystable}
\centering
\small\caption{Simulated MSEs as well as simulted variance and expectation of the estimated MSEs, under bivariate normal distribution, for different values of $\rho$, $n$ and $k$.\label{msere}}
\begin{tabular}{c c c | c c c c c |  c  c c c c }\hline\hline
Scheme & & &   &  RSS  &  & &   &       & DRSS &   &  &   \\
\hline
{$\rho$}&{$n$}&{$k$}&Simul. MSE &  Var($\hat{M}_{\gamma}$)& E($\hat{M}_{\gamma}$)&  Var($CV_{\gamma}$)& E($CV_{\gamma}$)&Simul. MSE &  Var($\hat{M}_{\gamma}$)& E($\hat{M}_{\gamma}$)&  Var($CV_{\gamma}$)& E($CV_{\gamma}$)\\
\hline
& & & & & &  & &  &  &  &  &                               \\
0.9& 15 & 3 & 0.0447 & 0.0293  & 0.0661 & 0.0001 & 0.0178 & 0.0405  & 0.0113 & 0.0455 & 0.0001 & 0.0175 \\
        &    & 5 & 0.0425 & 0.0039 & 0.0732 & 0.0008 & 0.0623 & 0.0422 & 0.0031 & 0.0683 & 0.0007 & 0.0612  \\
& & & & & &  & &   &  &    &  &                             \\
   & 30 & 3 &  0.0198 & 0.0038  &  0.0161 & 0.0001 & 0.0037 & 0.0172 & 0.0012& 0.0080 &  0.0001& 0.0036  \\
   &      & 5 &0.0160  & 0.0002 & 0.0117 & 0.0001 & 0.0103& 0.0151 &  0.0002 & 0.0115 & 0.0001 & 0.0104 \\
& & & & & &  & &   &  &   &  &                            \\
   & 45 & 3 & 0.0131 & 0.0005 & 0.0041 & 0.0001 & 0.0016 & 0.0114  & 0.0003 &  0.0024 & 0.0001 &  0.0015\\
   &      & 5 & 0.0099 & 0.0001 & 0.0040 & 0.0001 & 0.0039 & 0.0088 &0.0001 & 0.0041 & 0.0001 & 0.0041  \\
& & & & & &  & &  &  &  &  &                               \\
\hline
& & & & & &  & &  &  &  &  &                               \\
0.8& 15 & 3 & 0.0468 & 0.0439 & 0.0619 & 0.0001 & 0.0177 & 0.0429 & 0.0099 & 0.0441 & 0.0001 & 0.0177  \\
   &        & 5 & 0.0446  &0.0074 & 0.0814 & 0.0008 &  0.0442 & 0.0601 & 0.0018 & 0.0711 & 0.0008 &  0.0658  \\
& & & & & &  & & &  &    &  &                              \\
   & 30 & 3 & 0.0208  & 0.0077 & 0.0266 & 0.0001 & 0.0037 & 0.0184 & 0.0025 & 0.0103 & 0.0001 &  0.0036 \\
   &      & 5 & 0.0171  & 0.0007& 0.0145  & 0.0001 &  0.0102 & 0.0162 &0.0001 & 0.0113 & 0.0001 & 0.0106  \\
& & & & & &  & & &  & &  &                               \\
   & 45 & 3 & 0.0137  & 0.0019 & 0.0088 & 0.0001 &  0.0015 & 0.0122 & 0.0002& 0.0024& 0.0001 &  0.0016\\
   &      & 5 & 0.0106  &  0.0001 &  0.0052 & 0.0001 &  0.0039 & 0.0097 & 0.0001 &0.0041 & 0.0001 & 0.0041   \\
   & & & & & &  & &  &  &  &  &                               \\
\hline
\end{tabular}
\end{sidewaystable}
\subsection{Simulation study}

In order to compare the performance of $CV_{\gamma}$ in \eqref{Cg}
and $M_{\gamma}$ in \eqref{Mg}, two separate simulation studies were
conducted, each with 10,000 iterations. Since, computation of the
multiple integrals in $\alpha_i$ and $\beta_i$, $i=1,2$ takes a lot
of time, we have considered the simple case $p=2, p_1=p_2=1$ and
estimation of $H(X^{(1)})$,  under the bivariate normal distribution
with correlation parameter $\rho$, using the kernel in \eqref{knorm}
and the bandwidth in \eqref{rule}. The variable $X^{(2)}$ is
considered as the ranker variable and the computations are performed
for RSS and DRSS schemes. In the first simulation study, the values
of  $MSE(H_{n}^{\rm RSS}(X^{(1)}))$ were computed which are given in
Table \ref{msere}, for different values of $n$, $k$ and $\rho$, for
RSS and DRSS schemes. In the second simulation study, the variance
and expectation of $CV_{\gamma}$ and $M_{\gamma}$ were computed
which are given corresponding to each of the simulated
$MSE(H_{n}^{\rm RSS}(X^{(1)}))$ in Table \ref{msere}.

As one can see from Table \ref{msere}, the estimator $M_{\gamma}$ is less biased than  $CV_{\gamma}$, except for the case $k=5$ and $m=3$. As expected, the variance of  $CV_{\gamma}$ is less than that of $M_{\gamma}$. In the simulation study, it is found that the suitable values of $d_1$ for $H_{n}^{\rm RSS}(X^{(1)})$ are almost similar to those of
$CV_{\gamma}$ and $M_{\gamma}$.

\subsection{Data analysis}

We apply the results to a data set analyzed first by Platt et al.
(1988). The original data set consists of measurements made on 399
long-leaf pine (Pinus palustris) trees. We use a truncated version
of this data set, given in Chen et al. (2004), which contains the
diameter (in centimeters) at breast height ($X^{(1)}$) and the
height (in feet) ($X^{(2)}$) of 396 trees.

To examine the distributional properties of the proposed estimators for this real data set, we extract 10000
samples of size $n=mk=30$, with $k=3$ and $m=10$, using RSS, as well as samples of the same size using DRSS and SRS from this data set, by the diameter of the trees as the ranking variable.
 Since the population size is finite, the sampling is with replacement. Also, since the population does not have a density and the differential entropy is not defined, the true entropy of the population is estimated by
\[H_N(X)=\frac{1}{N}\sum_{i=1}^N\log f_N(X_i)I_S(X_i),\]
where $f_N(x)=\frac{1}{N\gamma}\sum_{i=1}^Nk_0\left(\frac{x-X_i}{\gamma}\right)$ and simillarly for $I(X^{(1)},X^{(2)})$. Using optimal bandwidth and kernel function for a simple random sample of size $N=396$, the full population estimates of the true parameters are $H_N(X^{(1)})=4.921,$  $I_N(X^{(1)},X^{(2)})=0.550$ and  ${\cal I}_N(X^{(1)},X^{(2)})=0.667$.

The simulated bias and MSE of the estimators were computed which are
given in Table \ref{simulreal}. As it can be observed from Table
\ref{simulreal}, the estimators of $I_N(X^{(1)},X^{(2)})$ and ${\cal
I}_N(X^{(1)},X^{(2)})$ based on DRSS have the smallest MSEs and
biases. Also, the estimators of $H_N(X^{(1)})$ has smallest MSE but
are more biased under RSS and DRSS schemes.

\begin{table}
  \centering
{  \caption{The simulated biases and MSEs of the entropy and the mutual information estimators for the tree data.}\label{simulreal}
\begin{tabular}{c |c c c }
  \hline\hline
Estimator & $H_n^{{\rm RSS}}(X^{(1)})$ & $H_n^{{\rm DRSS}}(X^{(1)})$ & $H_n^{{\rm SRS}}(X^{(1)})$ \\

\hline
Bias & -0.389 & -0.417 & 0.010  \\
MSE & 0.280 & 0.248 & 0.604  \\

\hline
Estimator& $\hat{I}^{\rm RSS}(X^{(1)},X^{(2)})$ & $\hat{I}^{\rm DRSS}(X^{(1)},X^{(2)})$ & $\hat{I}^{\rm SRS}(X^{(1)},X^{(2)})$\\
\hline
Bias & 1.111 & 1.069 & 1.953\\
MSE & 1.377 & 1.210 & 4.809 \\

\hline
Estimator& $\hat{\cal I}^{\rm RSS}(X^{(1)},X^{(2)})$ & $\hat{\cal I}^{\rm DRSS}(X^{(1)},X^{(2)})$ & $\hat{\cal I}^{\rm SRS}(X^{(1)},X^{(2)})$\\
\hline
Bias & 0.291 & 0.169 & 0.316\\
MSE & 0.085 & 0.084 & 0.100 \\
  \hline\hline
\end{tabular}}
\end{table}

\subsection{Application to a real data set}

In this section, we apply the proposed theoretical results to the problem of variable selection in regression estimation based on ranked set sampling (Yu and Lam, 1997).  We consider a real data set, consisting of measurements of different body fat percentage evaluations and various body circumference measurements for {$N=$252} men. This data set is taken from Penrose et al. (1985). Consider the following variables
\begin{description}
\item $Y$: Percent of body fat using Brozek's equation;
\item $X^{(1)}$: Abdomen circumference;
\item $X^{(2)}$: Weight;
\item $X^{(3)}$: Chest circumference;
\end{description}

Suppose, the regression estimator of the mean percent of body fat
using Brozek's equation is of interest. Three auxiliary variables
$X^{(1)}$, $X^{(2)}$ and $X^{(3)}$ are considered and suppose that
one is willing to choose the two most correlated auxiliary variables
with the variable of interest, $Y$. The classical correlation
criteria fail to measure the correlation between three variables.
Thus, the mutual information criterion is used as the variable
selection tool.

A double ranked set sample with $k=3$ and $m=10$, is extracted from
this data set, with $X^{(1)}$ as the ranking variable. This sample
is given in Table \ref{rssample}.

\begin{sidewaystable}
\centering \caption{Double ranked set sample extracted from the body fat data set. }\label{rssample}\small
\begin{tabular}{c | c | c c c c c c c c c c}
\hline\hline
Variable & &&&&&$i$&&&&\\
\hline
 & $j$ &1 & 2  & 3& 4  & 5 &  6  & 7 & 8 &  9 & 10\\
\cline{2-12}
                & 1        &  8.7 &10.9 &21.4&  9.2 & 5.0& 13.4&  0.0& 17.2& 14.4 &  8.8\\
 $Y_{[i]j}$& 2         & 25.8 &25.8 &15.5& 20.9 &25.9&  7.5& 20.5& 25.5&  6.5 & 25.9\\
                & 3        & 30.8 &16.5 &19.8& 27.1 &28.1& 16.5& 28.1& 26.2& 24.0 & 28.8 \\
 \cline{2-12}
  & &&&&&$i$&&&&\\
 \hline
& $j$ &1 & 2  & 3& 4  & 5 &  6  & 7 & 8 &  9 & 10\\
\cline{2-12}
                           & 1& 98.6  &99.2  &101.6 &109.3& 95.0& 86.0& 95.0& 91.6& 98.0&  82.9\\
$X^{(1)}_{[i]j}$    & 2& 86.1  &108.8 & 86.8 & 83.0& 95.0& 79.5& 79.1& 86.1& 90.8& 100.5\\
                          & 3& 104.3 &74.6  &78.2  &96.4 &92.1 &88.1 &89.1 &85.3 &97.5 & 79.4\\
 \cline{2-12}
  & &&&&&$i$&&&&\\
 \hline
& $j$ &1 & 2  & 3& 4  & 5 &  6  & 7 & 8 &  9 & 10\\
\cline{2-12}
                          & 1&177.00& 224.50& 203.25& 232.75& 178.25& 176.75& 198.5& 188.75& 202.25& 160.75\\
$X^{(2)}_{[i]j}$   & 2&166.75& 226.75& 159.25& 173.25& 198.50& 148.50& 168.0& 166.75& 172.75& 189.75\\
                          & 3&212.00& 131.50& 126.50& 187.75& 184.25& 157.75& 184.0& 188.15& 209.25& 140.50\\
 \cline{2-12}
  & &&&&&$i$&&&&\\
 \hline
& $j$ &1 & 2  & 3& 4  & 5 &  6  & 7 & 8 &  9 & 10\\
\cline{2-12}
                           & 1& 104.0& 113.2 &110.0 &117.5 & 99.7& 97.3 &106.5 &99.1 &109.2&  93.6\\
$X^{(3)}_{[i]j}$ & 2& 92.9 & 115.3 & 92.3 & 93.6 &106.5& 89.8 & 93.0 &92.9 & 99.1& 106.4\\
                           & 3& 106.6& 88.6  &88.8  &101.3 & 98.9& 97.5 &100.8 &96.6 &107.6&  91.2\\
 \hline\hline
\end{tabular}
\end{sidewaystable}

\begin{figure}
\centering
\includegraphics[scale=1]{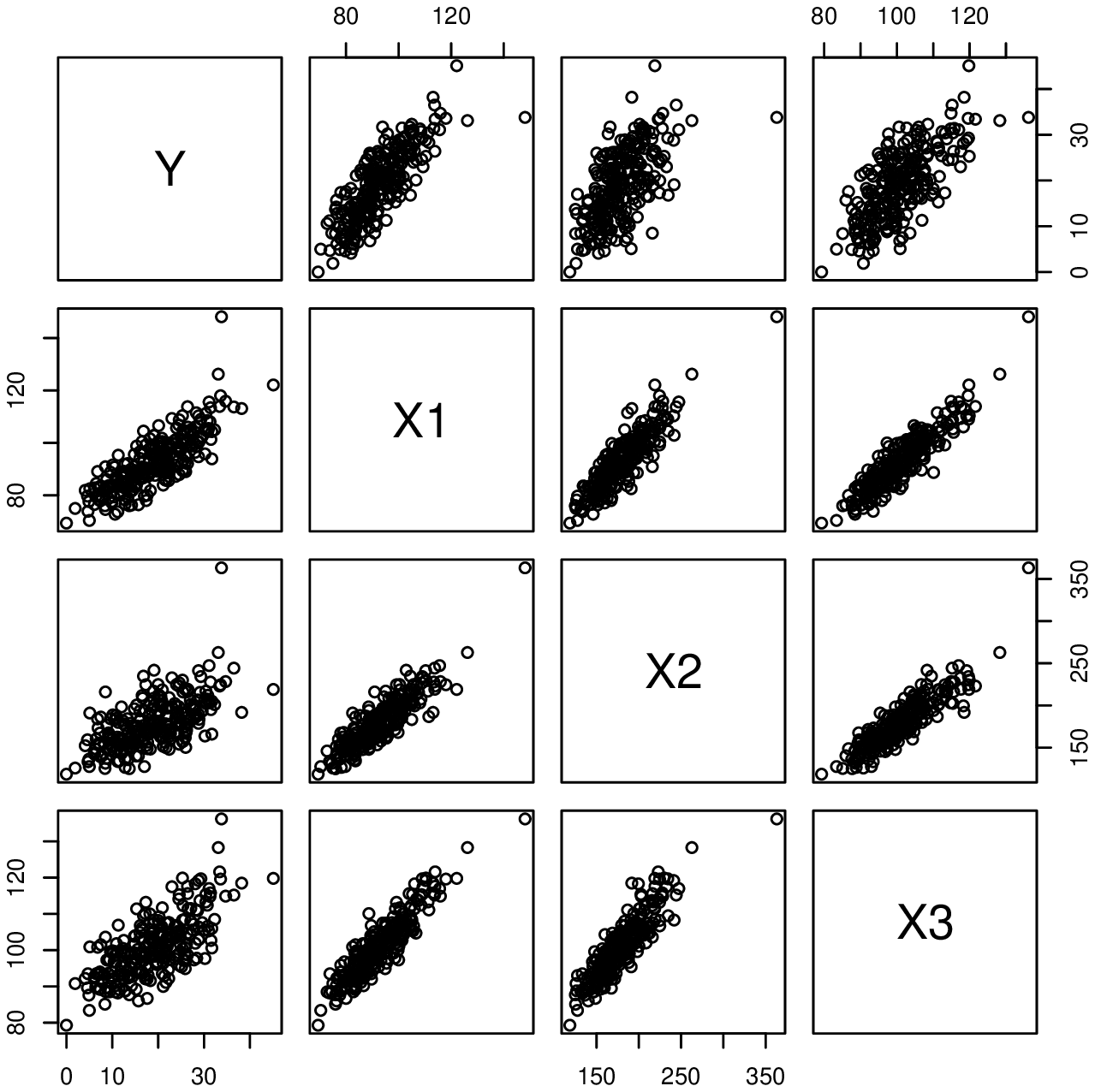}
\caption{Scatterplots for pairs of variables for the body fat data set}
\end{figure}

Based on this sample, the determinant of the correlation matrix of
the sample is obtained as
$$|R|=0.0077.$$
The simulation results under the multivariate normal distribution
with variance-covariance matrix $\Sigma$, satisfying
$|\Sigma|=0.01$, suggest using the bandwidth in \eqref{rule} with
$d_1=0.60$. The resulting estimators are

$${H}_n((Y))=3.599,\quad {H}_n((X^{(1)},X^{(2)},Y))=11.226,$$
$${H}_n((X^{(1)},X^{(2)}))=7.807,\quad {H}_n((X^{(1)},X^{(3)},Y))=10.464,$$
$${H}_n((X^{(1)},X^{(3)}))=6.894,\quad {H}_n((X^{(2)},X^{(3)},Y))=11.118,$$
$${H}_n((X^{(2)},X^{(3)}))=7.679,\quad \hat{I}((X^{(1)},X^{(2)},Y))=0.180,$$
$$\hat{\cal I}((X^{(1)},X^{(2)},Y))=0.302,\quad \hat{I}((X^{(1)},X^{(3)},Y))=0.029,$$
$$\hat{\cal I}((X^{(1)},X^{(3)},Y))=0.057,\quad \hat{I}((X^{(2)},X^{(3)},Y))=0.160,$$
and
$$\hat{\cal I}((X^{(2)},X^{(3)},Y))=0.274.$$

Since
$$\hat{\cal I}((X^{(1)},X^{(2)},Y))>\hat{\cal I}((X^{(2)},X^{(3)},Y))>\hat{\cal I}((X^{(1)},X^{(3)},Y)),$$
the variables $X^{(1)}$ and $X^{(2)}$ might be chosen out of the
three auxiliary variables, based on the sample given in Table
\ref{rssample}.

\section*{Acknowledgements}
The first author's research is supported in part by a grant BS-1393-1-02 from the Institute for Research in
Fundamental Sciences (IPM), Tehran, Iran.

\section*{Appendix}

\noindent{\bf Proof of Lemma 1}
The proof of part (i) is trivial. For the consistent ranked set sample, we have
\begin{align*}
\E\int\int a(x,y)\; {\rm d} U_{n}(x)\; {\rm d} U_{n}(y)&=n\E\int\int a(x,y)\; {\rm d} F_{n}^{\rm RSS}(x)\; {\rm d} F_{n}^{\rm RSS}(y)\\
&-n\int\int a(x,y)\; {\rm d} F_X(x)\; {\rm d} F_X(y)\\
&=n^{-1}\E\left\{\sum_{j=1}^m\left[\sum_{i=1}^k a(X_{[i]j},X_{[i]j})+\sum_{i_1\ne i_2}^{k}a(X_{[i_1]j},X_{[i_2]j})\right]\right.\\
&\left.+\sum_{j_1\ne j_2}^{m}\sum_{i_1=1}^k \sum_{i_1=1}^k a(X_{[i_1]j_1},X_{[i_2]j_2}) \right\}\\
&-n\int\int a(x,y)\; {\rm d} F_X(x)\; {\rm d} F_X(y)\\
&=\int a(x,x)\;{\rm d}F_X(x)-nm^{-1}\int \int a(x,y) \;{\rm d}F_X(x) \;{\rm d}F_X(y)\\
&+nm^{-1}\int \int a(x,y) \frac{1}{k^2}\sum_{i_1\ne i_2}^{k} f_{X_{[i_1]}}(x) f_{X_{[i_2]}}(y)\;{\rm d}x \;{\rm d}y\\
&=\int a(x,x)\; {\rm d} F_X(x)\\
&-\frac{1}{k}\sum_{i=1}^{k}\int\int a(x,y)f_{X_{[i]}}(x)f_{X_{[i]}}(y)\; {\rm d} x\; {\rm d} y.
\end{align*}

\noindent{\bf Proof of Theorem 1}
By the Taylor expansion of the integrands and using the fact that
$$f_n^{\rm RSS}(x)-\Delta_{\gamma}(x)=n^{-1/2}\int \frac{1}{\gamma^p}K_p\left(\frac{y-x}{\gamma}\right)\;{\rm d}U_n(y),$$
we can write
\begin{align*}
-H_{n}^{\rm RSS}(X)&=\int_{S}\log({f_{n}^{\rm RSS}(x)})\;{\rm d}F_{n}^{\rm RSS}(x)=\int_{S}(\log({f_{n}^{\rm RSS}(x)})-\log(\Delta_{\gamma}(x)))\;{\rm d}F_X(x)\\
&+n^{-1/2}\int_{S}(\log(f_{n}^{\rm RSS}(x))-\log(\Delta_{\gamma}(x)))\;{\rm d}U_{n}(x)\\
&+\int_{S}\log(\Delta_{\gamma}(x))\;{\rm d}F_X(x)+n^{-1/2}\int_{S}\log(\Delta_{\gamma}(x))\;{\rm d}U_{n}(x)\\
&=\int_{S}\frac{(f_{n}^{\rm RSS}(x)-\Delta_{\gamma}(x))}{\Delta_{\gamma}(x)}\;{\rm d}F_X(x)-\int_{S}\frac{(f_{n}^{\rm RSS}(x)-\Delta_{\gamma}(x))^2}{2\Delta_{\gamma}(x)^2}\;{\rm d}F_X(x)\\
&+n^{-1/2}\int_{S}\frac{(f_{n}^{\rm RSS}(x)-\Delta_{\gamma}(x))}{\Delta_{\gamma}(x)}\;{\rm d}U_{n}(x)\\
&+\int_{S}\log(\Delta_{\gamma}(x))\;{\rm d}F_X(x)+n^{-1/2}\int_{S}\log(\Delta_{\gamma}(x))\;{\rm d}U_{n}(x)+O(n^{-3/2}).
\end{align*}
Thus
\begin{align*}
H_{n}^{\rm RSS}(X)&=H_{\gamma}-n^{-1/2}\left[\int_{S}\int_{S}\frac{\frac{1}{\gamma^p}K_p\left(\frac{y-x}{\gamma}\right)}{\Delta_{\gamma}(y)}\;{\rm d}U_{n}(y)\;{\rm d}F_X(x)\right.\\
&-\left.\int_{S}\log(\Delta_{\gamma}(x))\;{\rm d}U_{n}(x)\right]\\
&-n^{-1}\left[\int_{S}\int_{S}\frac{\frac{1}{\gamma^p}K_p\left(\frac{y-x}{\gamma}\right)}{\Delta_{\gamma}(y)}\;{\rm d}U_{n}(y)\;{\rm d}U_{n}(x)\right.\\
&\left.+\int_{S}\int_{S}\int_{S}\frac{\frac{1}{\gamma^{2p}}K_p\left(\frac{y-x}{\gamma}\right)K_p\left(\frac{z-x}{\gamma}\right)}{2\Delta_{\gamma}(x)^2}\;{\rm d}U_{n}(y)\;{\rm d}U_{n}(z)\;{\rm d}F_X(x)\right]+O(n^{-3/2}).
\end{align*}
By applying Lemma \ref{l1}, the required result follows.
\hfill$\Box$

\section*{References}
\begin{enumerate}
\item  Al-Saleh, M.F., and Al-Kadiri, M. (2000). Double ranked set sampling. {\em Statistics \& Probability Letters}, {\bf 48}, 205-212.
\item Al-Saleh, M.F., and Al-Omari, A.I. (2002). Multistage ranked set sampling. {\em Journal of Statistical Planning and Inference}, {\bf 102}, 273-286.
\item  Chen, Z. (1999). Density estimation using ranked set sampling data. {\em Environmental and  Ecological Statistics,} {\bf 6},
135-146.
\item  Chen Z., Bai Z. and Sinha B. (2004). {\em Ranked set sampling: theory and applications.} Lecture Notes in Statistics.
Springer, New York.
\item Chen H, Stasny E.A. and Wolfe D.A. (2005). Ranked set sampling for efficient estimation of a population proportion.
{\em Statistics in Medicine}, {\bf 24}, 3319-3329.
\item  Cover, T. M. and Thomas, J. A. (2012). {\em Elements of Information Theory}. 2nd ed., Wiley, New York.
\item  Dell, T.R. and Clutter, J.L. (1972). Ranked set sampling theory with order statistics background. {\em Biometrics}, {\bf 28}, 545-555.
\item  Kvam, P.H. (2003). Ranked set sampling based on binary water quality data with covariates. {\em Journal of Agricultural, Biological, and Environmental Statistics}, {\bf 8}, 271–279.
\item  Halls L.K. and Dell T.R. (1966). Trial of ranked-set sampling for forage yields. {\em Forest Science}, {\bf 12}, 22–26.
\item  Deshpande J. V.,  Frey J. and Ozturk O. (2006).  Nonparametric ranked-set sampling confidence intervals
for quantiles of a finite population.   {\em Environmental and
Ecological Statistics}, {\bf 13}, 25-40.
\item Joe, H. (1989). Estimation of entropy and other functionals of a multivariate density. {\em Annals of the Institute of Statistical Mathematics}, {\bf 41}, 683-697.
\item Ozturk, O. (2011). Sampling from partially rank-ordered sets.  {\em Environmental and Ecological Statistics}, {\bf 18}, 757-779.
\item  Penrose, K.W., Nelson, A.G. and Fisher, A.G. (1985). Generalized body composition prediction equation
for men using simple measurement techniques. {\em Medicine and Science in Sports and Exercise}, {\bf 17},
189.
\item Platt, W.J., Evans G.W. and Rathbun S.L  (1988). The population-dynamics of a long lived conifer (Pinus palustris).
{\em American Naturalist}, {\bf 131}, 491-525.
\item McIntyre, G.A. (1952). A method of unbiased selective sampling using ranked sets. {\em Australian Journal of
Agricultural Research}, {\bf 3}, 385-390.
\item Shannon, C. E. (1968). A mathematical theory of communication. {\em Bell System Tech. J.}, {\bf 27}, 379-423.
\item  Stokes, S.L. (1980). Estimation of variance using judgement ordered ranked set samples, {\em Biometrics}, {\bf 36}, 35-42.
\item  Stokes, S.L. and Sager, T.W. (1988). Characterization of a ranked-set sample with application to estimating
distribution functions. {\em Journal of the American Statistical Association,}  {\bf 83}, 374-381.
\item Yu, P. L. H. and Lam, K. (1997). Regression Estimator in Ranked Set Sampling. {\em Biometrics}, {\bf 53 (3)}, 1070-1080.
\end{enumerate}
\end{document}